\documentclass[10pt,aps,prd,twocolumn,floats,floatfix,showpacs,superscriptaddress,nofootinbib]{revtex4-2}
\usepackage[utf8]{inputenc} 
\usepackage{graphicx,mathtools,amssymb,amsmath,amsthm,amsfonts,epsfig,epsf,bm,,multirow}
\usepackage[outdir=./]{epstopdf}
\usepackage[usenames]{color}
\usepackage{tensor}
\usepackage{csquotes}
\usepackage{mathrsfs}
\usepackage{autobreak}
\usepackage{physics}
\usepackage{cancel}
\usepackage[font=small,labelfont=bf, justification=Justified,format=plain]{caption}
\usepackage{slashed}
\usepackage{pifont}
\usepackage[dvipsnames]{xcolor}
\usepackage[colorlinks=true]{hyperref}
\usepackage{cleveref}

\def\be{\begin{equation}}
\def\ee{\end{equation}}
\def\bea{\begin{eqnarray}}
\def\eea{\end{eqnarray}}

\usepackage[normalem]{ulem}

\begin{document}

\title{A Lapse in the Cosmological Constant Problem}
\author{Justin Khoury}
\email{jkhoury@sas.upenn.edu}
\affiliation{\textsuperscript{*}Center for Particle Cosmology, Department of Physics and Astronomy, University of Pennsylvania, 209 South 33rd St, Philadelphia, PA 19104, USA}

\author{Benjamin Muntz}
\email{benjamin.muntz@nottingham.ac.uk}
\affiliation{Nottingham Centre of Gravity, University of Nottingham,
University Park, Nottingham NG7 2RD, United Kingdom}
\affiliation{School of Physics and Astronomy, University of Nottingham, University Park, Nottingham NG7 2RD, United Kingdom}

\author{Antonio Padilla}
\email{antonio.padilla@nottingham.ac.uk}
\affiliation{Nottingham Centre of Gravity, University of Nottingham,
University Park, Nottingham NG7 2RD, United Kingdom}
\affiliation{School of Physics and Astronomy, University of Nottingham, University Park, Nottingham NG7 2RD, United Kingdom}


\begin{abstract}
We present a new mechanism for addressing the cosmological constant problem based on global constraints arising from a lapse function in a higher-dimensional gravitational theory. Inspired by Hořava--Lifshitz gravity, we consider a 5d spacetime with anisotropic scaling along a compact extra dimension, while preserving Lorentz invariance in four dimensions. In the deep infrared limit, variation with respect to the lapse generates a global constraint on the 4d geometry, closely analogous to that of vacuum energy sequestering. Although the resulting effective gravitational equations differ from standard sequestering, radiative contributions to the Standard Model vacuum energy are nevertheless cancelled at all orders. 

\end{abstract}

\maketitle

\section{Introduction}
For a generic effective field theory (EFT) valid up to an ultraviolet cut-off scale
$\Lambda_{\rm cutoff}$, standard quantum field theory arguments imply that the vacuum
energy density receives radiative contributions of order~$\rho_{\rm vac} \sim \Lambda_{\rm cutoff}^4$.
In the absence of gravity this presents no inconsistency: a constant shift of the
Lagrangian has no effect on the equations of motion and is therefore physically
irrelevant.

The situation changes fundamentally once gravity is included. In General Relativity (GR),
vacuum energy gravitates universally and appears as an effective cosmological
constant. As a result,~$\rho_{\rm vac}$ is directly probed by large-scale cosmological
observations, which indicate an anomalously small value consistent with the observed
dark energy density,~$\rho_{\rm vac}^{\rm obs} \lesssim (\text{meV})^4$. For an EFT with a cut-off as low as the TeV scale, this corresponds to a discrepancy of
at least sixty orders of magnitude between the expected and observed vacuum energy
densities.

While it is always possible to cancel quantum contributions order by order by
introducing a finely tuned counterterm in the effective Lagrangian, this procedure
must be repeated at every loop order and for every threshold in the theory. The
resulting sensitivity of the gravitational dynamics to ultraviolet physics
represents an extreme violation of naturalness and signals a deep incompatibility
between standard EFT reasoning and the gravitational response to vacuum energy. For a range of reviews of the cosmological constant problem, see \cite{Weinberg:1988cp, Carroll:1991mt, Carroll:2000fy,Martin:2012bt,Polchinski:2006gy,Bousso:2007gp, Burgess:2013ara, Padilla:2015aaa, Bernardo:2022cck}.

One possibility is that the cosmological constant problem is signalling a structural failure in how vacuum energy gravitates. Modifying gravity to address cosmological problems at both short and long distances is nothing new \cite{Clifton:2011jh,Joyce:2014kja}. The difficulty with the cosmological constant is that it is a long wavelength source in the gravitational field equations. Modifying gravity at large distances is often in conflict with phenomenology, owing to the presence of new light degrees of freedom in the theory \cite{Niedermann:2017cel}. For example, these light fields can lead to fifth forces, spoiling precision tests of gravity. The standard response is to rely on so-called screening mechanisms that are often poorly understood. An interesting and perhaps unique exception to this rule is to consider the extreme case of a global modification of gravity, where the differences only apply to sources of infinite wavelength. This is a promising direction since vacuum energy is the infinite wavelength source of the gravitational field equations. All other sources like planets and stars have finite wavelength and should gravitate normally, enjoying the same observational successes as GR.

The prototypical example of this is \emph{vacuum energy sequestering} (VES), where new global degrees of freedom are added to the gravitational action \cite{Kaloper:2013zca, Kaloper:2014dqa, Kaloper:2014fca,Kaloper:2015jra,Kaloper:2016yfa,Kaloper:2016jsd,DAmico:2017ngr,Padilla:2018hvp,Coltman:2019mql,El-Menoufi:2019qva}. These new degrees of freedom act like Lagrange multipliers imposing global constraints. The result is that radiative
contributions to~$\rho_{\rm vac}$ do not gravitate. Localised sources trigger the same gravitational response as in GR.

However, the origin of the global degrees of freedom in VES is not so well understood. One heuristic possibility is motivated by wormhole configurations where fundamental constants can vary from one ``universe" to another \cite{Giddings:1987cg,Giddings:1988wv}. The resulting ensemble can be written as an integration over the value of the fundamental constants in the path integral \cite{Coleman:1988tj}, with the saddle points giving global constraints \cite{Padilla:2015aaa}. However, this interpretation is expected to exhibit the so-called wormhole catastrophe leading to a breakdown of semi-classical control and loss of predictability \cite{Fischler:1988ia,Polchinski:1989ae}.

On the other hand, the manifestly local formulation of VES \cite{Kaloper:2015jra} makes use of three-form fields and their corresponding four-form field strengths, extending the four-form formulations of unimodular gravity \cite{Henneaux:1989zc, Padilla:2014yea}. Such fields do not introduce any local dynamics but instead modify the global structure of the theory by enforcing global constraints. While this formulation opens up a possible pathway toward embedding VES in a UV-complete framework \cite{Padilla:2018hvp,El-Menoufi:2019qva}, a concrete realisation of such an embedding is still lacking.

In this paper, we explore a new route to VES-like global constraints, inspired by Hořava--Lifshitz gravity~\cite{Horava:2009uw,Horava:2009if}. Hořava--Lifshitz gravity is a proposal for quantum gravity in four spacetime dimensions in which Lorentz invariance is replaced in the ultraviolet by a preferred foliation and an anisotropic scaling between time and space. In the projectable version of the theory, the lapse function is restricted to depend only on the preferred time coordinate; a structure preserved by the underlying symmetries. As a result, variation of the action with respect to the lapse yields a global constraint on the spatial geometry.

Here we promote this idea to five spacetime dimensions by introducing a preferred spatial direction associated with a compact extra dimension. Lorentz invariance is preserved along four spacetime directions, with anisotropic scaling along the fifth dimension in the far infrared. We then perform an ADM decomposition adapted to this foliation of the 5d spacetime and take a projectable limit in which the lapse depends only on the preferred spatial coordinate. Analogous to the projectable version of Hořava-Lifshitz gravity, variation with respect to the lapse generates a global constraint along the foliation slices, which in our case corresponds to a 4d spacetime. We therefore obtain a global constraint on the 4d geometry, as in VES. Interestingly, however, the form of the constraint is subtly different from that of VES, leading to a distinct effective theory of gravity. The mechanism nevertheless cancels radiative contributions to the Standard Model vacuum energy and points to a one-parameter family of theories that generalises the global cancellations found in VES.

\section{Solving the cosmological constant problem with anisotropic scaling}
Lorentz invariance in four spacetime dimensions has been tested to extraordinary precision, with no violations observed at the level of one part in~$10^{21}$~\cite{Sanner:2018atx}. In theoretical models involving extra dimensions, it is often assumed that Lorentz symmetry extends to the full higher-dimensional theory. However, there is no direct experimental evidence for Lorentz invariance beyond the observed four dimensions (given there is also no evidence of extra dimensions!).

An intriguing and relatively unexplored possibility is that Lorentz invariance holds strictly in the observed 4d spacetime, while being explicitly broken along the extra dimensions via anisotropic scaling. This idea is reminiscent of Lifshitz field theories, originally introduced in the context of critical phenomena~\cite{Lifshitz:1941a,Lifshitz:1941b,Hornreich:1975}, in which some coordinates scale differently near a critical point. For example, consider a 5d scalar field~$\phi(x, y)$, where~$x^\mu$ denotes the four observed spacetime coordinates, and~$y$ is the coordinate of an extra dimension. A simple action exhibiting anisotropic scaling along the extra dimension takes the form
\begin{equation}
  S = \frac{1}{\kappa^2} \int \dd y\, \dd^4x\, \frac{1}{2} \phi \big( \Box_4 + \partial_y^{2z} \big) \phi\,,
\end{equation}
where~$\Box_4$ is the 4d d'Alembertian, and~$z$ is a positive integer controlling the scaling anisotropy. For~$z = 1$, the theory is Lorentz invariant in five dimensions. For general~$z$, the action transforms homogeneously under the anisotropic scaling
\begin{equation} \label{aniso}
  x^\mu \mapsto b^z x^\mu\,, \qquad y \mapsto b y\,.
\end{equation}
A field~$\phi$ is said to have scaling dimension~$[\phi]=\Delta$ if it transforms as~$\phi \mapsto b^\Delta \phi$. If we assume, without loss of generality, that the scalar field has vanishing scaling dimension,~$[\phi] = 0$, then the overall scaling of the action should be absorbed into a dimensionful coupling with scaling dimension~$[\kappa^2] = 2z + 1$.

Our goal here is to extend this idea to gravity. Specifically, we consider a 5d geometry where Lorentz invariance is preserved along the four spacetime directions, while allowing for anisotropic scaling in the extra dimension. To this end, we decompose the 5d geometry into four extended spacetime directions~$ x^\mu$ and a compact fifth dimension~$ y$.\footnote{Strictly speaking, neither the topology or size of the extra dimension will be important for the conclusion of this paper, as we consider the ultralocal limit that turns off all interactions along the extra dimension. In a follow-up paper~\cite{paper2}, however, we will show that in order to accommodate $z=1$ deformations, it is necessary to impose that it be small and compact with periodic boundary conditions.} In other words, we consider a foliation of the 5d spacetime by a family of 4d hypersurfaces~$ \{ \Sigma_y \}$, each at constant~$ y$, where~$ \Sigma_y$ represents a Lorentz-invariant slice of 4d spacetime.

To allow for anisotropic scaling in the extra dimension, we do not assume invariance under full 5d diffeomorphisms. Instead, we work with a restricted symmetry group corresponding to \emph{foliation-preserving diffeomorphisms}, both on-shell and off-shell. This idea is familiar from Hořava–Lifshitz gravity~\cite{Horava:2009uw, Horava:2009if}, where a preferred foliation of spacetime distinguishes temporal from spatial directions. In our case, the foliation separates the 4d spacetime coordinates~$ x^\mu$ from the extra dimension~$ y$.

With a view toward applying this framework to address the cosmological constant problem, we consider a theory analogous to the \emph{projectable} version of Hořava–Lifshitz gravity, restricting to foliations where the lapse function depends only on the transverse coordinate~$ y$. The 5d metric is then parametrised as
\begin{multline}
  \dd s_5^2 = N^2 \dd y^2 + g_{\mu\nu}\left(\dd x^\mu + N^\mu \dd y\right)\left(\dd x^\nu + N^\nu \dd y\right)\,,
\end{multline}
where~$ N(y)$ is the lapse function,~$ N^\mu(x, y)$ is the shift vector, and~$g_{\mu\nu}(x, y)$ is the induced 4d spacetime metric. This structure is preserved under foliation-preserving diffeomorphisms of the form
\begin{equation} \label{fdiffs}
  y \mapsto y - \xi(y)\,, \qquad x^\mu \mapsto x^\mu - \xi^\mu(x, y)\,,
\end{equation}
which maintain the integrity of the slicing by constant-$y$ hypersurfaces. Under the anisotropic scaling \eqref{aniso}, we assume that the lapse and 4d metric have vanishing scaling dimensions,~$[N] = [g_{\mu\nu}] = 0$. It is straightforward to check that the shift vector has dimension~${[N^\mu] = z - 1}$. In the following, we will introduce various~$p$-form fields,~$A_{\mu_1 \ldots \mu_p}$, whose scaling dimension we also take to vanish,~$[A_{\mu_1 \ldots \mu_p}]=0$.

We now consider a generalisation of 5d GR minimally coupled to matter, with~${z=0}$ anisotropic scaling. This corresponds to a deep infrared limit along the extra dimension, in which all~$y$-derivatives are absent. As a result, the action becomes ultralocal along the extra dimension: each slice at fixed 
$y$ evolves independently, with no coupling between different slices. Phenomenologically, this does not influence any local observational signatures whatsoever---every slice can essentially be thought of as an independent universe. It is however the modification of the \emph{global} spacetime structure which allows the imposing of global constraints relevant to solving the cosmological constant problem.

The action we consider is given by
\begin{equation}\label{eq:action}
\begin{split}
  S &= \frac{1}{2\kappa^2} \int \dd y \, N \int_{\Sigma_y} \dd^4x \Big(\sqrt{-g}\, R - F_4 \wedge \star_4 F_4\Big)  \\
  &\qquad + S_m[N, g_{\mu\nu}, \Psi]+ \text{boundary terms}\,,
\end{split}
\end{equation}
where indices are raised and lowered with the 4d metric~$g_{\mu\nu}$, whose Ricci scalar is~$R$. The matter sector includes the Standard Model fields,~$\Psi(x, y)$, smeared along the extra dimension. For~$z=0$ scaling, the matter action can be decomposed as
\begin{equation}
  S_m[N, g_{\mu\nu}, \Psi] = \int \dd y \, N \int_{\Sigma_y} \dd^4x \sqrt{-g} \, \mathcal{L}_{\text{SM}}(g_{\mu\nu}, \Psi)\,,
\end{equation}
where~$\mathcal{L}_{\text{SM}}$ is the usual Standard Model Lagrangian with minimal coupling to the spacetime metric~$g_{\mu\nu}$ (understood precisely as its  quantum effective action with graviton loops neglected). We also introduce a 3-form field 
\begin{equation}
  A_3(x, y) \equiv \frac{1}{3!} A_{\mu_1 \mu_2 \mu_3}(x, y) \, \theta^{\mu_1} \wedge \theta^{\mu_2} \wedge \theta^{\mu_3}\,,
\end{equation}
where the co-frames are given by~$\theta^\mu = \dd x^\mu + N^\mu \, \dd y$. Note that the components of~$A_3$ transform covariantly under the foliation-preserving diffeomorphisms \eqref{fdiffs}. The corresponding field strength is given by~$F_4 = \dd_4 A_3$, the exterior derivative of the three-form~$A_3$ taken along the 4d spacetime slices. The Hodge star~$\star_4$ is defined with respect to these 4d slices. For~$z=0$, the coupling~$\kappa^2$ has scaling dimension~$[\kappa^2]=1$.

The boundary terms in \eqref{eq:action} lie along possible boundaries of the 4d spacetime. These include the Gibbons–Hawking-York (GHY) term~\cite{Gibbons:1976ue} 
\begin{equation}
  \frac{1}{\kappa^2}\int \dd y \, N \int_{\partial \Sigma_y} \dd^3 \xi \sqrt{-\gamma}\,  K \,,
\end{equation}
and the Duncan–Jensen term~\cite{Duncan:1989ug}
\begin{equation} \label{DJz=0}
  \frac{a}{\kappa^2} \int \dd y \, N \int_{\partial \Sigma_y} A_3 \wedge \star_4 F_4 \,,
\end{equation}
where the boundary surface~$\partial \Sigma_y$ has induced line element~$\gamma_{ij}(\xi, y) \dd\xi^i \dd \xi^j$ and extrinsic curvature~$K_{ ij}$. The Duncan-Jensen term allows us to interpolate between Dirichlet ($a=0$) or Neumann ($a=1$) boundary conditions for the 3-form field on this surface. The GHY term is typically included to ensure Dirichlet boundary conditions on the metric. However, let us examine the boundary terms that arise when we vary the gravitational part of the action 
\begin{equation}
  \int \dd y\, N \left[\int_{\Sigma_y} \dd^4 x\, \sqrt{-g} \frac{1}{2 \kappa^2} R + \int_{\partial \Sigma_y} \dd^3 \xi\, \sqrt{-\gamma}\, \frac{1}{\kappa^2} K \right]
\end{equation}
with respect to both the metric and the lapse. This yields the boundary variation 
\begin{equation} \label{boundaryvariation}
  \int_{\partial \Sigma_y} \dd^3 \xi\, \sqrt{-\gamma}\, \frac{1}{\kappa^2} \left[ -\frac{1}{2} \big(K^{ij} - K \gamma^{ij}\big) \delta \gamma_{ij} + K \frac{\delta N}{N} \right]\,.
\end{equation}
As expected, this term vanishes under Dirichlet boundary conditions on both the metric and the lapse at~$\partial \Sigma_y$. Notwithstanding, since the lapse depends only on~$y$, imposing Dirichlet conditions would freeze all fluctuations in the bulk. This is too restrictive for our purposes: as we will see, the equations of motion obtained from varying the lapse play a crucial role in addressing the cosmological constant problem. To ensure the boundary variation \eqref{boundaryvariation} vanishes, we must impose the following mixed boundary on the lapse and induced metric,
\begin{equation} \label{bc}
N\delta \gamma_{ij} = -\gamma_{ij}\delta N \,.
\end{equation}
This scenario is reminiscent of the boundary considerations discussed in~\cite{Kaloper:2016yfa}. It is equivalent to imposing Dirichlet boundary conditions on the metric upon transforming to the Einstein frame,~$ \tilde{g}_{\mu\nu} = N g_{\mu\nu}$, where the prefactor in~\eqref{eq:action} associated with the lapse function is absorbed into the rescaled metric on the spacetime hypersurfaces.

Having endured a slightly tedious boundary distraction, we are now ready to compute the equations of motion that emerge from bulk variation of the corresponding fields. The equation for the 3-form implies
\begin{equation}
\star_4 F_4=Q(y)\,,
\end{equation}
where~$Q$ is a function of~$y$ only. The gravitational equations of motion include variations with respect to the lapse
\begin{equation} \label{Neqn}
  \int \dd^4 x \sqrt{-g} \left[ 
  \frac{1}{2}R_4 +\left(\frac12-a\right) Q^2  +\kappa^2 \mathcal{L}_\text{SM} \right] =0
\end{equation}
and the metric
\begin{equation} \label{geqn}
G_{\mu\nu}=-\frac{Q^2}{2} g_{\mu\nu} +\kappa^2 T^\text{(SM)}_{\mu\nu}\,.
\end{equation}
Here~$T^\text{(SM)}_{\mu\nu}=-\frac{2}{\sqrt{-g}}\frac{\delta }{\delta g^{\mu\nu}}\int_{\Sigma_y} \dd^4x \sqrt{-g} \, \mathcal{L}_{\text{SM}}$ is the energy--momentum tensor for the matter sector. Note that the shift vector drops out of \eqref{eq:action}. It will return when we consider deformations to~$z=1$ or higher~\cite{paper2}. 

The key point to note is that the lapse equation~\eqref{Neqn} does not impose a local condition on the 4d spacetime geometry. Rather, it enforces a global constraint. This arises directly from working off-shell in the projectable limit of the 5d geometry, where the lapse function depends only on the slicing coordinate~$y$. Note that the Duncan–Jensen boundary term \eqref{DJz=0} contributes to this global constraint, whereas the choice of mixed boundary condition \eqref{bc} ensures that the GHY term does not. 

The presence of a global constraint on the 4d geometry is strongly reminiscent of VES \cite{Kaloper:2013zca, Kaloper:2014dqa, Kaloper:2014fca,Kaloper:2015jra,Kaloper:2016yfa,Kaloper:2016jsd,DAmico:2017ngr,Padilla:2018hvp,Coltman:2019mql,El-Menoufi:2019qva}. In that framework, global constraints arise from auxiliary fields that act to decouple vacuum energy from gravity. The underlying setup here is quite different, relying on an extra dimension rather than global variables. The effective gravitational equations obtained after integrating out the global degrees of freedom also differ. Nevertheless, we will now show how vacuum energy is still decoupled from the gravitational dynamics.

To this end, we solve \eqref{Neqn} for~$Q(y)$ and substitute the result into \eqref{geqn}. After some trivial rearranging, we arrive at the following effective gravitational equation, 
\begin{equation}
G_{\mu\nu}=\kappa^2\left(T^\text{(SM)}_{\mu\nu}-\frac{1}{2(3-2a)}\left\langle T^\text{(SM)} -2 \mathcal{L}_\text{SM} \right\rangle g_{\mu\nu} \right)\,,
\end{equation}
where~$T^\text{(SM)}$ is the trace of the energy-momentum tensor, and the angled brackets denote {\it spacetime} averages, 
\begin{equation}
  \expval{\mathcal{O}} \equiv \frac{\int \dd^4x\sqrt{-g}\ \mathcal{O}}{\int \dd^4x\sqrt{-g}}\,.
\end{equation}
This quantity is obviously well defined for finite spacetime volume, and even in infinite spacetime whenever the operator $\mathcal{O}$ has finite support. For further discussions on this definition, we refer to the literature on VES \cite{Kaloper:2013zca, Kaloper:2014dqa, Kaloper:2014fca,Kaloper:2015jra,Kaloper:2016yfa,Kaloper:2016jsd,DAmico:2017ngr,Padilla:2018hvp,Coltman:2019mql,El-Menoufi:2019qva}.

If we impose Neumann boundary conditions on the 3-form ($a=1)$, the effective gravity equation reduces to the following form
\begin{equation} \label{geffeqna=1}
G_{\mu\nu}=\kappa^2\left(T^\text{(SM)}_{\mu\nu}-\frac{1}{2}\left\langle T^\text{(SM)} -2 \mathcal{L}_\text{SM} \right\rangle g_{\mu\nu} \right)\,,
\end{equation}
or equivalently
\begin{multline}
G_{\mu\nu}=\kappa^2\bigg(  g_{\mu\nu} \big({\cal L_\text{SM}} - \langle{\cal L_\text{SM}}\rangle \big) \\   - 2 \frac{\delta {\cal L_\text{SM}}}{\delta g^{\mu\nu}} + \left\langle g^{\rho\sigma} \frac{\delta {\cal L_\text{SM}}}{\delta g^{\rho\sigma}}\right\rangle g_{\mu\nu} \bigg)\,, 
\end{multline}
which is invariant under~${\cal L}_\text{SM} \mapsto {\cal L}_\text{SM} + {\rm const}$. Therefore, this equation has the remarkable property that the Standard Model vacuum energy drops out. To see this explicitly, we write the Lagrangian as~${\mathcal{L}_\text{SM}=-V_\text{vac}+\mathcal{L}_\text{local}}$, where~$V_\text{vac}$ corresponds to the renormalised vacuum energy, and~$\mathcal{L}_\text{local}$ describes local excitations about the vacuum. This yields~$T^\text{(SM)}_{\mu\nu}=-V_\text{vac} g_{\mu\nu} +T^\text{(local)}_{\mu\nu}$, where~$T^\text{(local)}_{\mu\nu}=-\frac{2}{\sqrt{-g}}\frac{\delta }{\delta g^{\mu\nu}}\int_{\Sigma_y} \dd^4x \sqrt{-g} \, \mathcal{L}_{\text{local}}$ is the energy--momentum tensor for local matter excitations about the vacuum. Plugging this result into \eqref{geffeqna=1}, we obtain
\begin{equation} 
G_{\mu\nu}=-\Lambda_\text{eff}+\kappa^2T^\text{(local)}_{\mu\nu}\,,
\end{equation}
where 
\begin{equation}
 \Lambda_\text{eff}= \frac{\kappa^2}{2}\left\langle T^\text{(local)} -2\mathcal{L}_\text{local} \right\rangle \,.
\end{equation}
Naturalness arguments suggest that the Standard Model vacuum energy,~$V_\text{vac}$, should be at least of order~$(\text{TeV})^4$. In GR, such a large contribution to the gravitational field equations would be disastrous for cosmology without extreme fine-tuning. In our case, this problem is avoided:~$V_\text{vac}$ drops out of the dynamics entirely. This does not mean there is no cosmological constant---it exists, given by~$\Lambda_\text{eff}$---but it depends only on local excitations of the matter field \emph{around} the vacuum and is insensitive to the radiative instabilities afflicting the vacuum energy.

\section{Discussion}
We have presented a novel set-up involving anisotropic extra dimensions that leads to global geometric constraints along 4d slices and the cancellation of the Standard Model vacuum energy. These cancellations are strongly reminiscent of those seen in VES, but they are not identical: consider the one-parameter family of theories whose dynamics generalises~\eqref{geffeqna=1},
\begin{equation} \label{geffeqn-gen}
G_{\mu\nu}=\kappa^2\left(T^\text{(SM)}_{\mu\nu}-\frac{1}{4}\left\langle (1+\alpha)T^\text{(SM)} -4\alpha \mathcal{L}_\text{SM} \right\rangle g_{\mu\nu} \right)\,.
\end{equation}
For any value of~$\alpha$, the vacuum energy drops out of the equation. VES corresponds to~$\alpha=0$, while our new framework described by~\eqref{geffeqna=1} corresponds to~$\alpha = 1$.

The set-up presented here corresponds to taking the deep infrared limit along the extra dimension. In a companion paper \cite{paper2}, we will explore ultraviolet deformations of the theory obtained by including operators with~$z=1$ scaling behaviour. Such deformations do not spoil the cancellation mechanism presented here. This is perhaps not surprising, since vacuum energy is an infinite-wavelength source whose gravitational effects are controlled by the far infrared. We will also consider what happens when full five dimensional diffeomorphisms are restored using the St\"uckelberg trick. This will reveal the crucial role played by boundary conditions on the St\"uckelberg fields in ensuring cancellations of vacuum energy are robust even when the effects of phase transitions are taken into account.

There are some similarities between our set-up and a related proposal involving global constraints by Carroll and Remmen~\cite{Carroll:2017gqo}. The effective gravity equation derived in~\cite{Carroll:2017gqo} is indeed identical to~\eqref{geffeqna=1}. However, a key difference lies in the nature of the global variable. In~\cite{Carroll:2017gqo}, Planck’s constant~$\hbar$ is promoted to a global variable, whose variation yields a global constraint analogous to~\eqref{Neqn}. As explained in~\cite{DAmico:2017ngr}, this mechanism works only at tree level: quantum corrections to the vacuum energy appear with different powers of~$\hbar$, modifying the constraint and spoiling the cancellation. In contrast, our proposal identifies the global variable with the projected 5d lapse function. Quantum corrections to the vacuum energy do not introduce new powers of the lapse, thanks to the underlying foliation-preserving diffeomorphism symmetry. As a result, the cancellation mechanism remains valid beyond tree level. 

Finally, while we have focussed on radiative corrections to vacuum energy from matter loops, we expect that our model will also be protected when graviton loops are included. This is because {\it  all} vacuum energy corrections, regardless of whether or not they include graviton loops,  contribute to the quantum effective action with the same power of the lapse. Again, this follows automatically from the foliation preserving diffeomorphisms. We will discuss this issue further in our forthcoming companion paper \cite{paper2}.

\acknowledgements
We thank the Corfu Summer Institute where this work was initiated. AP acknowledges support from STFC Consolidated Grant nos. ST/V005596/1 and ST/X000672/1. The work of JK is supported in part by the DOE (HEP) Award No. DE-SC0013528. For the purpose of open access, the authors have applied a CC BY public copyright licence to any Author Accepted Manuscript version arising. No new data were created during this study. 

\bibliography{ref}

\end{document}